# Realizing high-quality, ultra-large momentum states using semiconductor hyperbolic metamaterials

Salvatore Campione,[1,2,*] Sheng Liu,[1,2] Ting S. Luk,[1,2] and Michael B. Sinclair[2,#]

[1]Center for Integrated Nanotechnologies (CINT), Sandia National Laboratories, P.O. Box 5800, Albuquerque, NM 87185, USA
[2]Sandia National Laboratories, P.O. Box 5800, Albuquerque, NM 87185, USA
[*]sncampi@sandia.gov      [#]mbsincl@sandia.gov

**Abstract:** We employ both the effective medium approximation (EMA) and Bloch theory to compare the dispersion properties of semiconductor hyperbolic metamaterials (SHMs) at mid-infrared frequencies and metallic hyperbolic metamaterials (MHMs) at visible frequencies. This analysis reveals the conditions under which the EMA can be safely applied for both MHMs and SHMs. We find that the combination of precise nanoscale layering and the longer infrared operating wavelengths puts the SHMs well within the effective medium limit and, in contrast to MHMs, allows the attainment of very high photon momentum states. In addition, SHMs allow for new phenomena such as ultrafast creation of the hyperbolic manifold through optical pumping. In particular, we examine the possibility of achieving ultrafast topological transitions through optical pumping which can photo-dope appropriately designed quantum wells on the femtosecond time scale.

## 1. Introduction

Layered media [1] provide great flexibility for tailoring the optical properties of surfaces and are utilized in a wide variety of optical devices. Recently, a new and fascinating class of layered media called hyperbolic metamaterials (HMs) has been introduced [2]. They are usually formed by alternating subwavelength layers of positive and negative permittivity materials (other implementations do exist, e.g. using nanowires [3, 4] or graphene [5, 6]), and in the effective medium limit behave as uniaxial materials with hyperbolic isofrequency wavevector surfaces [7]. As a result, these materials are characterized by extremely large (infinite in the effective medium limit) densities of states which can greatly enhance spontaneous emission [8, 9], enhance near-field thermal energy transfer [10], and lead to enhanced absorption processes [11, 12].

The degree to which these remarkable properties appear depends upon the range of photon momenta over which the isofrequency surface remains hyperbolic. This, in turn, depends directly upon the underlying multilayer architecture of the HMs: at sufficiently large photon momentum, the out-plane momentum approaches the Brillouin zone edge and the effective medium description ceases to be valid. Thus, the momentum range of hyperbolic behavior increases as the layer thicknesses are decreased relative to the operating wavelength.

At optical and visible frequencies HMs are generally implemented as metal/dielectric stacks [13]. Due to the relatively large thicknesses in comparison to the operating wavelength, these metallic hyperbolic metamaterials (MHMs) have a limited band of wavevectors in which they actually support hyperbolic dispersion [11, 14, 15]. One might think that this drawback can be easily overcome by utilizing thinner layers or operating in the mid-infrared. However, the preparation of extremely thin, continuous metal films is difficult, and the film thickness probably can't be reduced by much more than a factor of two without substantially increasing losses. Furthermore, at mid-infrared frequencies many multilayer periods are required to achieve a metamaterial of appreciable thickness, which would be difficult for MHMs. Of course, future improvements of metal deposition techniques might make large numbers of thinner metal layers possible.

At mid-infrared frequencies, highly doped semiconductor materials behave like metals. As a result, semiconductor hyperbolic metamaterials (SHMs) can be fabricated using alternate layers of doped and undoped semiconductor materials [16-18]. While MHMs have shown properties that may lead to practical applications, SHMs offer opportunities for improved properties and can also allow for new phenomena such as *ultrafast creation of the hyperbolic manifold through optical pumping*. Furthermore, semiconductor layers can easily be fabricated with nanometer scale layer thicknesses which, along with the mid-infrared (~10 µm) operating wavelengths, places SHMs well within the effective medium regime.

Thus, it is expected that SHMs will exhibit an increased momentum range over which they exhibit hyperbolic dispersion when compared to MHMs at visible frequencies [18, 19].

Our first objective in this paper is to employ both the effective medium approximation (EMA) and Bloch theory [20, 21] to compare the dispersion properties of SHMs at mid-infrared frequencies and MHMs at visible frequencies. This analysis will allow us to clearly establish the conditions under which the EMA can be applied safely for both MHMs and SHMs. Next, we will examine the possibility of transiently inducing or destroying hyperbolic behavior through optical pumping. In this way, we can optically induce electrons to populate or leave the doped layers [22, 23] of the SHMs and achieve ultrafast topological transitions.

## 2. Dispersion properties of SHMs: Effective medium approximation and Bloch theory

According to the EMA, a HM made of alternating doped/undoped semiconductor materials or metal/dielectric materials (with relative permittivities $\varepsilon_d$ and $\varepsilon_m$ for dielectric and metallic layers, respectively) can be described using a homogeneous, uniaxial permittivity tensor of the kind $\underline{\varepsilon}_{HM} = \varepsilon_t \hat{t} + \varepsilon_l \hat{l}$, where $\varepsilon_t = \frac{\varepsilon_m d_m + \varepsilon_d d_d}{d_m + d_d}$ is the transverse permittivity along the transverse direction $\hat{t}$ and $\varepsilon_l = \left(\frac{\varepsilon_m^{-1} d_m + \varepsilon_d^{-1} d_d}{d_m + d_d}\right)^{-1}$ is the longitudinal permittivity along the longitudinal direction $\hat{l}$. In these expressions $d_d$ and $d_m$ represent the dielectric and metallic layer thicknesses. For such a uniaxial material, the dispersion relations for ordinary and extraordinary waves satisfy the following equations:

$$\text{Ordinary}: k_t^2 + k_l^2 = k_0^2 \varepsilon_t \qquad \text{Extraordinary}: \frac{k_t^2}{\varepsilon_l} + \frac{k_l^2}{\varepsilon_t} = k_0^2 \qquad (1)$$

with $k_t$ and $k_l$ the transverse and longitudinal wavenumbers, respectively, and $k_0 = \omega/c$ the free space wavenumber, with $\omega$ the angular frequency and $c$ the speed of light.

According to Bloch theory, the dispersion relations for ordinary and extraordinary waves are obtained by computing $k_l$ from the expression

$$e^{ik_l d} = \frac{(A+D) \pm \sqrt{(A+D)^2 - 4}}{2} \qquad (2)$$

with $d = d_m + d_d$ (the period of the HM), and by choosing the "$\pm$" sign such that the wave is decaying [20, 21]. In Eq. (2), $A$ and $D$ are the diagonal elements of the unit cell transfer matrix which is obtained by multiplying the matrices describing the metallic and dielectric layers:

$$\begin{pmatrix} A & B \\ C & D \end{pmatrix} = \begin{pmatrix} A_m & B_m \\ C_m & D_m \end{pmatrix} \times \begin{pmatrix} A_d & B_d \\ C_d & D_d \end{pmatrix} = $$
$$\begin{pmatrix} \cos(k_{lm} d_m) & -iZ_m \sin(k_{lm} d_m) \\ -i\sin(k_{lm} d_m)/Z_m & \cos(k_{lm} d_m) \end{pmatrix} \times \begin{pmatrix} \cos(k_{ld} d_d) & -iZ_m \sin(k_{ld} d_d) \\ -i\sin(k_{ld} d_d)/Z_d & \cos(k_{ld} d_d) \end{pmatrix} \qquad (3)$$

where the subscripts $m$ and $d$ refer to the metallic and dielectric layers, respectively. In this expression $k_{lm,d} = \sqrt{k_0^2 \varepsilon_{m,d} - k_t^2}$ are the longitudinal wavenumbers in the metallic and dielectric layers; $Z_{m,d} = k_{lm,d}/(\omega \varepsilon_{m,d} \varepsilon_0)$ for TM polarization and $Z_{m,d} = \omega \mu_0/k_{lm,d}$ for TE polarization; and $\varepsilon_0$ and $\mu_0$ are the absolute permittivity and permeability of free space, respectively. [The monochromatic time harmonic convention, $\exp(-i\omega t)$, is implicitly assumed.] This formulation is valid for semiconductor layers that are sufficiently thick such that the electronic energy quantization can be ignored and their permittivities can be modeled using an isotropic Drude [24, 25] model. Thinner layers should be modeled as uniaxial layers, with Drude transverse permittivities, and either dielectric or Drude longitudinal permittivities; and the Bloch theory described above should be modified to accommodate uniaxial layers.

We now analyze the following two structures (see Fig. 1): an experimentally achievable design for an SHM using doped InAs ($2 \times 10^{19}$ cm$^{-3}$) and undoped GaSb layers [26]; and a MHM with silver and glass. For the SHM case, the (relative) permittivities $\varepsilon_m$ and $\varepsilon_d$ are described by a Drude model

$$\varepsilon_{m,d} = \varepsilon_\infty \left(1 - \frac{\omega_p^2}{\omega^2 + i\omega\gamma}\right) \quad (4)$$

where $\varepsilon_\infty$ is given by the Sellmeier's equation for InAs and GaSb, $\omega_p = 458.94 \times 10^{12}$ rad/s is the plasma angular frequency which is proportional to the doping density and the effective mass for InAs ($\omega_p$ is equal to zero in GaSb), and $\gamma = 6.56 \times 10^{12}$ rad/s is the damping rate which is proportional to the mobility and the effective mass for InAs. For the MHM case, the dielectric permittivity is $\varepsilon_d = 2.25$, while the metallic permittivity is modeled as in Eq. (4) using $\varepsilon_\infty = 3.5$, $\omega_p = 7.299 \times 10^{15}$ rad/s, and $\gamma = 151.72 \times 10^{12}$ rad/s. The use of alternative models for the permittivity of silver would not affect the conclusions drawn in this paper. We assume for the moment that the thicknesses of metallic and dielectric layers of both types of HMs to be $d_m = 20$ nm and $d_d = 40$ nm, respectively. These layer thicknesses are easily achieved for the semiconductors and are also comparable to the thicknesses used in many of the experimental implementations of MHMs at visible wavelengths [2, 13, 27-30].

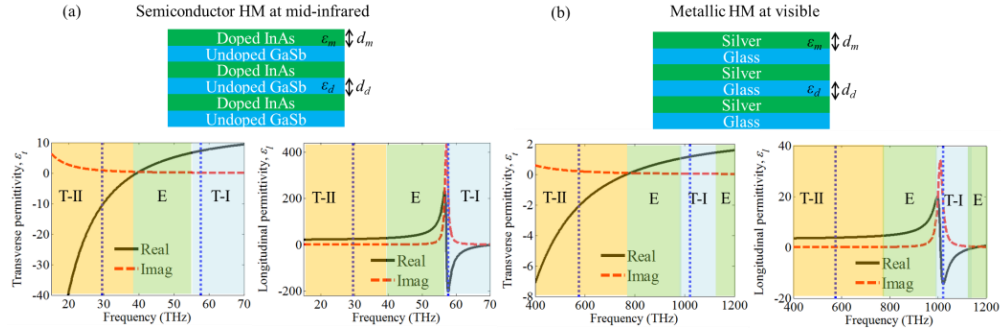

Fig. 1. (a) Transverse ($\varepsilon_t$) and longitudinal ($\varepsilon_l$) dielectric constants of a SHM made of alternating doped InAs and undoped GaSb layers as in the top schematic. (b) As in (a), for a MHM made of alternating silver and glass layers as in the top schematic. Here, $d_m = 20$ nm and $d_d = 40$ nm. The blue dotted lines indicate the frequencies adopted for the results in subsequent figures. The colored boxes indicate a type II hyperbolic dispersion (T-II) region, an elliptic dispersion (E) region, and a type I hyperbolic dispersion (T-I) region.

Figure 1 shows the frequency dependence of the transverse and longitudinal permittivities ($\varepsilon_t$ and $\varepsilon_l$) for both types of HMs obtained using the effective medium equations. For both HMs, $\text{Re}(\varepsilon_t) < 0$ and $\text{Re}(\varepsilon_l) > 0$ at low frequencies, leading to a frequency band of type II hyperbolic dispersion [31]. The transverse permittivity increases with increasing frequency and eventually becomes positive, leading to a frequency band of elliptic dispersion. As the frequency is further increased, a resonance occurs in the longitudinal permittivity, above which $\text{Re}(\varepsilon_l) < 0$, leading to a band of type I hyperbolic dispersion [31]. Finally, at high frequencies both $\varepsilon_t$ and $\varepsilon_l$ are positive and elliptic dispersion is obtained. We also observe that the real parts of $\varepsilon_t$ and $\varepsilon_l$ assume larger values in the SHM case.

The $k_l - k_t$ dispersion diagrams obtained using both EMA and Bloch theory for the two HMs are shown in Fig. 2. These diagrams were calculated for two frequencies. The first frequency is chosen to lie in the frequency band of type II hyperbolic dispersion and corresponds to 75% of the frequency where $\text{Re}(\varepsilon_t) = 0$. This choice results in 29.55 THz for the SHM and 576 THz for the MHM. The second frequency is chosen to be in the frequency band of type I hyperbolic dispersion, in particular at the frequency where the real part of the longitudinal permittivity has a global minimum (57.7 for the SHM and

1023 THz for the MHM). For the SHM at mid-infrared frequencies, we observe good agreement between EMA and Bloch theory up to transverse wavenumbers of about $k_t/k_0 = 20$, while for the MHM at visible frequencies the two theories agree only up to a transverse wavenumbers of about $k_t/k_0 < 2$. This is simply a result of the larger ratio of the operating wavelength to the period for the SHM. The normalized Brillouin zone boundaries $\text{Re}(k_l) = \pm\pi/(k_0 d)$ (shown as dotted blue lines in Fig. 2(b)) are reached in the case of the MHM.

As also mentioned in [11], above the $k_t$ where the value of $\text{Re}(k_l)$ reaches the Brillouin zone, the attenuation constant $\text{Im}(k_l)$ increases dramatically, resulting in a highly evanescent HM spectrum. The spectral range for which EMA and Bloch theory are in agreement can be further investigated using the quality factor $Q = |\text{Re}(k_l)/\text{Im}(k_l)|$ [16, 32, 33]. Figure 3 shows the quality factors obtained using both the EMA and Bloch theories for both types of HM. In the region of type I hyperbolic behavior, the quality factor decreases rapidly with increasing transverse wavenumber $k_t$ for both the EMA and Bloch theories. In contrast, in the region of type II hyperbolic behavior of the semiconductor metamaterial (Fig. 3(a)), the Bloch quality factor is ~20 up to large values of the transverse wavenumber $k_t$, while the EMA quality factor remains constant over the entire range shown. This means that in the SHM system the electromagnetic wave is underdamped and will oscillate multiple times before dying out. We note that if we were to use InGaAs/AlInAs instead of InAs/GaSb, the quality factor would be about $Q_{\max} \approx 8$, about three times smaller, because the mobility of InGaAs is four times smaller than that of InAs (2500 cm$^2$/(V-s) versus 10,000 cm$^2$/(V-s)), leading to higher material losses for InGaAs. For the MHM, the quality factor in the region of type II behavior reaches a maximum value of ~15 at $k_t/k_0 = 2.5$, but rapidly decreases for higher $k_t$ values according to Bloch theory. This rapid decrease is due to the onset of Brillouin zone-edge behavior.

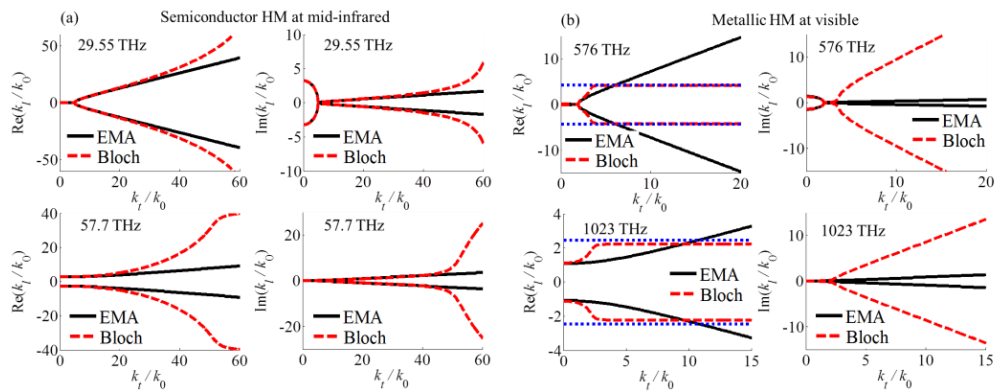

Fig. 2. (a) $k_l - k_t$ dispersion diagrams of a SHM as in Fig. 1 for two frequencies as indicated in the panels. (b) As in (a), for a MHM as in Fig. 1. In (b), the normalized Brillouin zone boundaries $\text{Re}(k_l) = \pm\pi/(k_0 d)$ are shown as dotted blue lines.

One can define the spectral bandwidth $B$ as the width of the waveband where the system is underdamped, i.e. it exhibits $Q > 1$, as qualitatively indicated by the arrows in Fig. 3. Qualitatively, this bandwidth is larger for the SHM than for the MHM, although this figure makes it difficult to quantify. Analogous to the gain-frequency product of an operational amplifier [34], we define the $Qk_t/k_0$ product of a hyperbolic metamaterial (see Fig. 4a). Assuming a given value for $Q$, the larger the value of $Qk_t/k_0$, the larger the band of transverse momenta that can be used for applications requiring HMs, e.g. radiative energy transport [19]. The $Qk_t/k_0$ product for both types of HMs is shown in Fig. 4(b) and Fig. 4(c). For the SHM, the product reaches a maximum value of 310 in the type I region and 942 in the type II region. For the MHM, a value of 31 is achieved in the type I region and 41 in the type II region. This significant difference between the two systems is due to the larger $Q$ values and the larger range of $k_t$ values where the EMA is applicable for the SHM case. The maximum of the $Qk_t/k_0$ product denotes the maximum transverse momentum for which the system has a large quality factor (i.e. it is underdamped). In the type II region, the maximum of the $Qk_t/k_0$ product occurs at $k_t/k_0 \approx 45$ for the SHM, and only $k_t/k_0 \approx 3$ for the MHM. These maxima are indicated by green dots in Fig. 4(b) and Fig. 4(c). These results imply that the SHM under analysis at mid-infrared frequencies exhibits better hyperbolic properties than the MHM under consideration at visible frequencies.

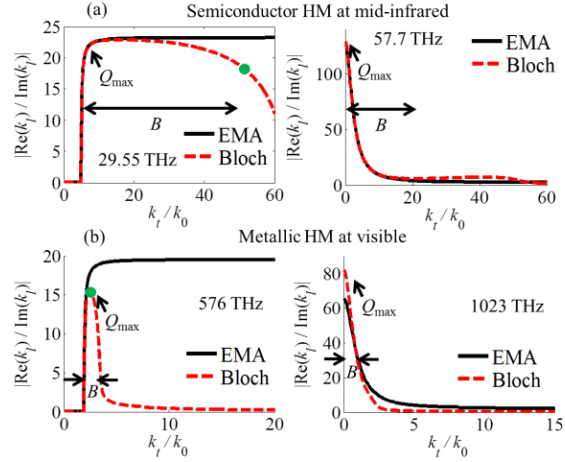

Fig. 3. (a) Quality factor defined as $Q = |\text{Re}(k_l)/\text{Im}(k_l)|$ relative to the dispersion diagram result in Fig. 2 for a SHM as in Fig. 1 for two frequencies as indicated in the panels. (b) As in (a), for a MHM as in Fig. 1. We qualitatively indicate the peak of the quality factor $Q_{\max}$ and the spectral bandwidth $B$ where the system is underdamped. The green dots indicate the spectral position for T-II that will be discussed in Fig. 4.

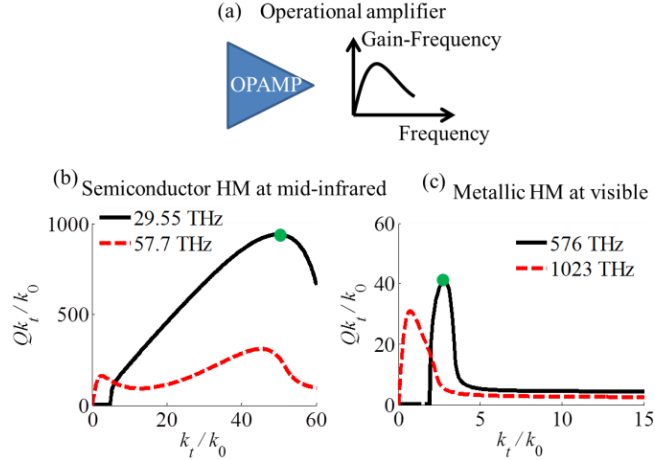

Fig. 4. (a) Gain-frequency product of an operational amplifier, analogous to the $Qk_t/k_0$ product of a hyperbolic metamaterial. (b) Product $Qk_t/k_0$ relative to the result in Fig. 3 for a SHM as in Fig. 1 for two frequencies as indicated in the panels. (c) As in (b), for a MHM as in Fig. 1. The values for the SHM at mid-infrared are always larger than for the MHM at visible.

## 3. Accuracy of the effective medium approximation for SHMs

We now concentrate only on the SHM and investigate the accuracy of the EMA as the metallic and dielectric layer thicknesses are increased. As mentioned earlier, the Brillouin zone edge defines the maximum range of propagating $k_l$ waves, and thus limits the maximum supported $k_t$. We keep the same metal filling factors as previously and consider the same frequencies analyzed in Figs. 2-4. Figure 5 shows the quality factors achieved assuming the following layer thicknesses: $d_m = 20\,\text{nm}$ and $d_d = 40\,\text{nm}$ [

$(\pi/d)/k_0 = 85$ and $(\pi/d)/k_0 = 43$ at 29.55 and 57.7 THz, respectively], $d_m = 30$ nm and $d_d = 60$ nm, $d_m = 40$ nm and $d_d = 80$ nm, and $d_m = 50$ nm and $d_d = 100$ nm [$(\pi/d)/k_0 = 34$ and $(\pi/d)/k_0 = 17$ at 29.55 and 57.7 THz, respectively]. This figure shows that as the size of the unit cell increases, the range of transverse momenta for which the EMA applies decreases. In particular, while for a period of 60 nm the EMA applies up to $k_t/k_0 = 20$, the EMA only applies up to $k_t/k_0 = 10$ for a period of 150 nm, showing the rapid breakdown of EMA for increasing unit cell period. This is partly due to the fact that the wavelength *inside the semiconductor material* is much smaller than its free-space value. For example, at 29.55 THz the wavelength is $\lambda_l = \text{Re}(2\pi/k_l) \approx 13d$ when $d = 60$ nm and $\lambda_l \approx 4d$ when $d = 150$ nm when evaluated at $k_t/k_0 = 20$.

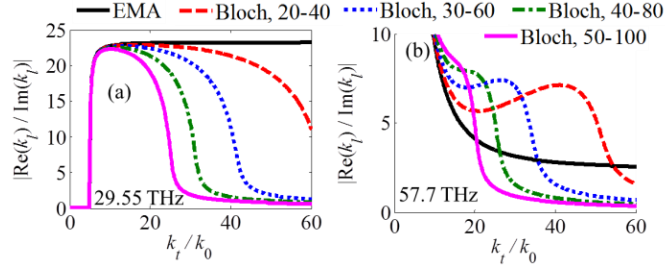

Fig. 5. (a) Quality factor for a SHM as in Fig. 1 computed at (a) 29.55 THz and (b) 57.7 THz, assuming the following layer thicknesses: $d_m = 20$ nm and $d_d = 40$ nm, $d_m = 30$ nm and $d_d = 60$ nm, $d_m = 40$ nm and $d_d = 80$ nm, and $d_m = 50$ nm and $d_d = 100$ nm. The breakdown of EMA for increasing unit cell period is evident.

## 4. Ultrafast topological transitions in SHMs

We finally show that SHMs enable ultrafast topological transitions between elliptical and hyperbolic dispersion. Consider again the structure used in Fig. 2(a), but now with lower doping for the InAs layer ($1 \times 10^{18}$ cm$^{-3}$). Using an ultrafast optical pump, we can rapidly produce a large density ($\sim 2 \times 10^{19}$ cm$^{-3}$) of photo carriers that will be trapped within the quantum wells (i.e., the doped layers). In this manner the quantum wells are rapidly switched from dielectric to metallic behavior at the frequencies of interest, and the superlattice structure will switch from elliptic to hyperbolic dispersion. Figure 6 shows the dispersion relations for the un-pumped and photo-pumped conditions for the T-II hyperbolic region. A transition from elliptical to hyperbolic dispersion is clearly observed, and the $Qk_t/k_0$ product again reaches a maximum value of 942. After initial pumping, the superlattice will remain in the hyperbolic state for a brief period (~1 ns) before relaxation processes deplete the photo-carriers. A similar behavior can also be achieved for the T-I hyperbolic region. This intriguing result suggests that transient excitation of SHMs might enable optical gating and modulation for many applications, such as imaging, lensing, and near-field energy transfer, as well as all-optical switching. We stress that experimental verification of such a photo-pumping scheme will be extremely challenging, but should be achievable.

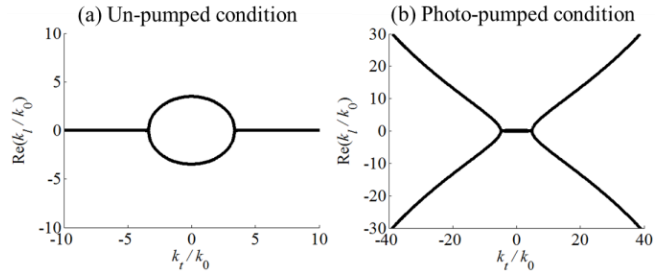

Fig. 6. $k_l - k_t$ dispersion diagrams (real part only) of a SHM as in Fig. 1 at 29.55 THz in (a) un-pumped ($1 \times 10^{18}$ cm$^{-3}$) and (b) photo-pumped ($2 \times 10^{19}$ cm$^{-3}$) conditions. The dispersion changes from an ellipse to a hyperbola.

## 5. Conclusion

In conclusion, we have shown that semiconductor hyperbolic metamaterials at mid-infrared frequencies can be accurately described by the effective medium approximation for a wide range of photon momenta. This result is largely achieved due to the large ratio of the wavelength to the superlattice period which postpones the onset of band-edge cutoff effects until large values of the transverse wavenumber. For the metal structures analyzed at visible frequencies, the hyperbolic behavior is rapidly quenched due to band-edge cutoff. Of course, the situation for MHMs can be improved by utilizing thinner layers or operating at mid-infrared. However, the preparation of extremely thin, continuous metal films is difficult and the film thickness probably can't be reduced by much more than a factor of two without increasing losses. Furthermore, at mid-infrared frequencies many multilayer periods are required to achieve a metamaterial of appreciable thickness, which would be difficult for MHMs. We also showed that SHMs can achieve large quality factors, even larger than ones for MHMs at visible frequencies. In addition, we proposed that, through the use of semiconductor superlattice materials, one could optically induce electrons to populate or leave the doped layers [22, 23], thereby enabling topological transitions on ultrafast timescales.

**Acknowledgments**

The authors acknowledge fruitful discussions with Caner Guclu, University of California Irvine. This work was supported by the U.S. Department of Energy, Office of Basic Energy Sciences, Division of Materials Sciences and Engineering and performed, in part, at the Center for Integrated Nanotechnologies, an Office of Science User Facility operated for the U.S. Department of Energy (DOE) Office of Science. Sandia National Laboratories is a multi-program laboratory managed and operated by Sandia Corporation, a wholly owned subsidiary of Lockheed Martin Corporation, for the U.S. Department of Energy's National Nuclear Security Administration under contract DE-AC04-94AL85000.